\documentclass[aps,prd,superscriptaddress,twoside,twocolumn,nofootinbib,showpacs]{revtex4-1}
\usepackage{amsmath,amssymb}
\usepackage{mathtools}
\usepackage{bbold}
\usepackage[usenames]{color}
\usepackage{braket}
\usepackage{graphicx,bm}
\usepackage{slashed}
\usepackage{ulem}
\usepackage{slashbox}
\usepackage[usenames]{color}
\renewcommand\sout{\bgroup \color{red} \ULdepth=-.5ex \ULset}
\begin{document}

\title{\boldmath Complex phase structure of the meson-baryon $T$-matrix}

\author{Shahab Razavi}%
\email{srazavi@uga.edu}
\affiliation{Department of Physics and Astronomy, The University of Georgia,
Athens, GA 30602, USA}

\author{K. Nakayama}%
\email{nakayama@uga.edu}
\affiliation{Department of Physics and Astronomy, The University of Georgia,
Athens, GA 30602, USA}
%\affiliation{Institut f\"ur Kernphysik and Center for Hadron Physics,
%Forschungszentrum J\"ulich, 52425 J\"ulich, Germany}

\date{\today}

\begin{abstract}
The full complex phase structure of the meson-baryon reaction amplitude in coupled channels approach is investigated, including also the photon-baryon channel. The result may be viewed as a generalization of the well-known Watson's theorem. Furthermore, the complex phase structure is exhibited for the pole and nonpole parts of the reaction amplitude in such a way that it will serve as a convenient common starting point for constructing models with different levels of approximation, in particular, for building isobar models where the basic properties of the $S$-matrix can be maintained. Such models should be useful, especially, in coupled multichannel calculations, where a large amount of experimental data are considered in resonance analyses, a situation encountered in modern baryon spectroscopy. 
In particular, it is shown that the unitarity of the pole part of the $T$-matrix arises automatically from the dressing mechanism inherent in the basic scattering equation.
This implies that no separate conditions are required for making this part of the resonance amplitude unitary as it has been done in some of the existing isobar models.

\end{abstract}


\maketitle

\section{Introduction}\label{sec:Intro}

Baryon spectroscopy is an important part of the study of non-perturbative regime of QCD. So far, most of the known baryon resonances have been identified in $\pi N$ scattering experiments. With recent progresses in this field, it is clear that a reliable resonance identification and extraction of its properties from experimental spectra requires a consistent analysis of many independent reaction processes. Coupled channels approach is the tool of choice for this task. Indeed, reaction theories based on coupled channels approach have been developed at various degrees of sophistication. Nowadays, such analyses in baryon spectroscopy involve coupled multichannel calculations analyzing a large amount of experimental data in various meson production channels. These data are being accumulated at major facilities world wide, especially, in photoproduction reactions. %
The most sophisticated coupled channels approach is that of Dynamical Coupled Channels (DCC) developed over many years %\cite{*[{}][{ and references therein.}]MSL07,DLMS07,Diaz08,Diaz09,Kamano09,KNLS13,KNS14,KNLS15,*[{Erratum: }][{}]KNLS15Erratum,DORING2009170,DHHKMR11,HDHHHKMN12,Rönchen2013,Rönchen2014,*[{Erratum: }][{}]Rönchen2014Erratum,RDHHMN15,RDM18}.
\cite{MSL07,DLMS07,Diaz08,Diaz09,Kamano09,KNLS13,KNS14,KNLS15,DORING2009170,DHHKMR11,HDHHHKMN12,DR2013,DR2014,RDHHMN15,RDM18}.
These calculations are quite involved and it is customary to make some sort of approximations in order to keep such calculations numerically more manageable. 
A common such approximation is the $K$-matrix 
%\cite{Kmatrix1,Kmatrix2,Kmatrix3} 
approach and its variations employed by some of the resonance analyses groups %\cite{BnGa10,BnGa11,ABKNST12a,ABKNST12b,BnGa13,BnGa17,PhysRevC.71.055206,*[{Erratum: }][{}]PhysRevC.71.055206Erratum,Giessen05b,Giessen13.11,Giessen16,Kent03,Kent12,Kent13,Zagreb10,GWU06,PhysRevC.74.045205,PhysRevC.86.035202,PhysRevC.72.025205, 10.1143/PTPS.186.216}. 
\cite{BnGa10,BnGa11,ABKNST12a,ABKNST12b,BnGa13,BnGa17,PhysRevC.71.055206,Giessen05b,Giessen13.11,Giessen16,Kent03,Kent12,Kent13,Zagreb10,GWU06,PhysRevC.74.045205,PhysRevC.86.035202,PhysRevC.72.025205, 10.1143/PTPS.186.216}. 
A nice feature of the $K$-matrix approach is that it reduces the original scattering equation to an algebraic equation while preserving unitarity of the $S$-matrix. This feature enables incorporating a large amount of experimental data in coupled multichannel analyses.  

A particular variation of the $K$-matrix approach is the so-called isobar models, where the reaction amplitude is decomposed into a resonance and a background contribution. Basically, they should correspond to the pole and nonpole parts of the  $T$-matrix amplitude. The background amplitude is usually parametrized by some smooth functions of energy while the resonance amplitude is parametrized by Breit-Wigner forms. Isobar models are practical and very economical in performing numerically demanding calculations, and are often used in resonance analyses based on coupled channels calculations and also dealing with a large amount of experimental data. Despite being simple, isobar models still capture many interesting properties of the resonances.
One issue that arises in these models is that unitarity is usually violated. There are many efforts to unitarize isobar models \cite{PhysRev.113.338,doi:10.1146/annurev.ns.13.120163.002011,M66,GM67,O74a,%AB78,
BB79,L88,DHKT99,DKT07,TDKV11,PhysRevC.67.015209}.  There, the resonance and the background amplitudes are unitarized separately and independently. This leads to a quite involved constraint on the resonance amplitude in particular. 
One of the unitary isobar models used intensively in the analyses of both the photo- and electro-production reactions is that of Mainz group \cite{DHKT99,DKT07,TDKV11,Tiator2018}. In their approach, the unitarization of the background amplitude is done by solving the scattering equation for that amplitude. For the resonance pole amplitude, based on Ref.~\cite{O74a}, it introduces complex resonance coupling constants which are constrained by imposing the unitarity condition independent from the background amplitude. Recently the Mainz group has updated its etaMAID isobar model \cite{etaMAID} by introducing a constant complex phase to each of their resonance amplitudes. Note that, in principle, the complex phase is an energy-dependent function containing proper threshold behaviors. Complex phases in the resonance coupling constants have been also introduced in the study of  hadronic reactions (see, e.g., Ref.~\cite{LMDHN19}).

In the present work, we exhibit the full complex phase structure of the meson-baryon $T$-matrix reaction amplitude in coupled channels approach. 
To this end, we first expose the complex phase structure of the full reaction amplitude written in terms of the $K$-matrix and the so-called \textit{generalized} Watson's factor. The result may be considered  as a generalization of the well-known Watson's theorem in photoproduction \cite{Watson54}.   
This helps us to expose, in a second step, the full complex phase structure of the pole and nonpole parts of the reaction amplitude which serves as a common staring point for introducing approximations to the reaction amplitude with varying degrees of sophistication. The resulting form of the reaction amplitude is such that the fundamental properties of the $S$-matrix, such as unitarity and/or analyticity, can be maintained straightforwardly in different approximations. In particular, we show how the unitarity of the pole part of the $T$-matrix arises automatically from the dressing mechanism inherent in the basic $T$-matrix equation, and that, no separate conditions are required for making this part of the resonance amplitude unitary as it has been done in some of the existing isobar models. 

This paper is organized as follows. In Sec.~\ref{sec:Notation}, we introduce the notation used throughout this work for the sake of conciseness. In Sec.~\ref{sec:PS1}, we derive the full phase structure of the meson-baryon reaction amplitude which is essentially a generalization of the Watson's theorem.
Based on this, the complex phase structure of the pole and nonpole parts of the reaction amplitude is derived in Sec.~\ref{sec:PS2}.  In Sec.~\ref{sec:PS3}, the phase structure of the photoproduction amplitude in one-photon approximation is derived. 
In Sec.~\ref{sec:UAIM}, possible levels of approximation to the full reaction amplitude are briefly discussed.
A summary is given in Sec.~\ref{sec:SUM}. 
For completeness, the phase-shift parametrizations of the $T$- and $K$-matrices as well as of the generalized Watson's factor are given in Appendix~\ref{ap:PSparam}. Since the decomposition of the $T$-matrix into the pole and nonpole parts plays a central role in the present work, this decomposition is derived in Appendix~\ref{ap:C} for both the meson-baryon and photoproduction reaction processes.
%The latter is done starting from the gauge-invariant amplitude in one-photon approximation based on the field theoretic approach
  Appendix~\ref{ap:TRC} contains the explicit form of the dressed resonance propagator in the case of the two resonance coupling.

\section{Notation}\label{sec:Notation}

Before starting the derivation of the complex phase structure of the meson-baryon reaction amplitude, a  remark on the notation used in the present work is in order.

The two-body reaction amplitude $T$ obeys, in general, the Lippmann-Schwinger-type scattering equation (also referred to as the $T$-matrix equation) 
\begin{equation}
T = V + VGT \ ,
\label{eq:B0}
\end{equation}
where $V$ denotes the driving potential kernel, irreducible with respect to the ``two-particle cut" \cite{SlpeterBethe}, and  $G$ stands for the two-body propagator. Note that the above equation is an integral equation for operators in abstract space.   

In momentum space, and in the coupled channels approach, the above equation becomes
\footnote{The relativistic generalization of the scattering equation given by Eq.~(\ref{eq:B1}) - the so called Bethe-Salpeter equation \cite{SlpeterBethe} - involving a four-dimensional momentum integration, may be reduced to a three-dimensional integral equation of the form given by Eq.~(\ref{eq:B1}) in such a way to maintain Lorentz covariance and elastic unitarity of the original reaction amplitude \cite{CJ88}. This means that Lorentz covariance can be also maintained in three-dimensional scattering equations, along with the other basic properties of the $S$-matrix, such as unitarity and analyticity.}
\begin{align}
T_{\nu'\nu}&(\vec{q}\,', \vec{q}; E)  =  V_{\nu'\nu}(\vec{q}\,', \vec{q}) \nonumber \\
& + \sum_\lambda \int d^3 q'' V_{\nu'\lambda}(\vec{q}\,', \vec{q}\,'') G_\lambda(\vec{q}\,''^2, E) T_{\lambda\nu}(\vec{q}\,'', \vec{q}; E) \ .
\label{eq:B1}
\end{align}
Here, $\vec{q}\,'$, $\vec{q}$, and $\vec{q}\,''$, denote the final, initial and intermediate two-particle relative momenta, respectively.  $E$ stands for the total energy of the system. The indices $\nu'$, $\nu$, and $\lambda$ stand for the final, initial and intermediate two-particle channels.

The reaction amplitude given by Eq.~(\ref{eq:B1}) can be expanded in partial-waves as
\begin{align}
& \bra{S'M_{S'}} T_{\nu'\nu}(\vec{q}\,', \vec{q}; E) \ket{S M_S}   = \nonumber \\
& =  \sum i^{L - L'}  (S' M_{S'} L' M_{L'} | J M_J) (S M_S L M_L | J M_J)  \nonumber \\
& \qquad \ \ \times T^{JI S'S}_{\nu'\nu\, L'L} (q', q; E)  Y_{L' M_{L'}}(\hat{q}') Y^{*}_{L {M_L}}(\hat{q}) \hat{P}_I  \ ,
\label{eq:B2}
\end{align}
where $S$, $L$, and $J$ denote the spin, orbital angular momentum and total angular momentum, respectively of the two-body  initial state, while $M_S$, $M_L$ and $M_J$ stand for the corresponding projection quantum numbers. The primed quantities refer to the corresponding quantum numbers of the final two-body state.  $\hat{P}_I$ stands for the isospin projection operator which projects the two-body state onto the total isospin state $I$. $Y_{lm_l}(\hat{p})$ stands for the usual spherical harmonic function. Here, the argument $\hat{p}$ is a short-hand notation for the polar ($\theta$) and azimuthal ($\phi$) angles, i.e., $\hat{p}=(\theta_p, \phi_p)$. 
The geometrical factor $(j_1 m_1 j_2 m_2| j_3 m_3)$ is the usual $SU(2)$ Clebsch-Gordan coefficient.
The summation in the above equation is over all the quantum numbers appearing on the right-hand side and not specified on the left-hand side of the equation.

The partial-wave amplitude $T^{JIS'S}_{\nu'\nu\, L'L} (q', q; E)$ in Eq.~(\ref{eq:B2}) can be extracted by inverting that equation. We have
\begin{align}
& T^{JI S'S}_{\nu'\nu\, L'L} (q', q; E)  =  \sum i^{L' - L}  \left(\frac{8\pi^2}{2J+1}\right) \left(\frac{2L+1}{4\pi}\right)^{\frac12}  \nonumber \\
& \times (S' M_{S'} L' M_{L'} | J M_J) (S M_S L 0 | J M_J)\  \hat{P}_I \nonumber \\
& \times \int^{+1}_{-1} d(\cos\theta_{q'})  \bra{S'M_{S'}} T_{\nu'\nu}(\vec{q}\,', \vec{q}; E) \ket{S M_S} \nonumber \\
&  \qquad\qquad  \times Y^*_{L' M_{L'}}(\theta_{q'}, 0)   \ ,
\label{eq:B3}
\end{align}
where, without loss of generality, the initial relative momentum $\vec{q}$ is chosen along the $+z$-axis and the final relative momentum $\vec{q}\,'$ in the $x$-$z$ plane. Similarly to Eq. (\ref{eq:B2}), the summation in the above equation is over all the quantum numbers appearing on the right-hand side and not specified on the left-hand side of the equation.

Inserting Eq.~(\ref{eq:B2}) into (\ref{eq:B1}), yields the scattering equation for the partial-wave amplitude
\begin{align}
T^{JI S'S}_{\nu'\nu\, L'L}(q', q; E) & =  V^{JI S'S}_{\nu'\nu\, L'L}(q', q) \nonumber \\
& + \sum_{S'',L'',\, \lambda}  \int^{\infty}_0  d q'' q''^2 V^{JI S'S''}_{\nu'\lambda\, L'L''}(q', q'') \nonumber \\
& \qquad\qquad 
\times G_\lambda(q''^2, E) T^{JI S''S}_{\lambda\nu\, L''L}(q'', q; E) \ .
\label{eq:B4}
\end{align}

In the present work, we use the notation
\begin{equation}
T_{\alpha'\alpha} = V_{\alpha'\alpha}  + \sum_\beta V_{\alpha'\beta} G_\beta T_{\beta\alpha} 
\label{eq:B5}
\end{equation}
to denote either Eq.~(\ref{eq:B1}) or (\ref{eq:B4}) for the sake of conciseness. 
 Accordingly, if the above equation is to represent Eq.~(\ref{eq:B1}), the indices $\alpha'$, $\alpha$ and $\beta$ in the above equation stand for the two-particle channel of the final, initial and intermediate states, respectively, and the summation over $\beta$ is to be understood as the summation over the intermediate two-particle channels. On the other hand, if the above equation is to represent Eq.~(\ref{eq:B4}), then, the indices $\alpha'$, $\alpha$ and $\beta$ specify, in addition to the two-particle channel of the final, initial and intermediate states, respectively, also the corresponding two-body partial-wave states.  Note also that the reference to the two-particle relative momentum is completely suppressed in the present notation, including its integration over the intermediate states. 
 
The notation explained above is used throughout the present paper. In particular, the main result of this work, given by Eqs. (\ref{eq:F28}) and (\ref{eq:PF28}), can be interpreted as given either in plane-wave or in partial-wave basis.

\section{Phase structure of the two-body $T$-matrix amplitude}\label{sec:PS1}

To expose the phase structure of the two-body reaction amplitude, it is convenient to express the $T$-matrix in terms of the $K$-matrix. We start with the $T$-matrix scattering equation 
\begin{equation}
T = V  + TGV = V + VGT \ ,
\label{eq:A1}
\end{equation}
where the two-body propagator $G$ can, in general, be decomposed into the real and imaginary parts
\begin{equation}
G = G^R - iG^I \ .
\label{eq:A2}
\end{equation}
In fact, the propagator involving stable particles is of the form ($\epsilon \to 0$)
\begin{equation}
G = \frac{1}{E - H_0 + i\epsilon} = {\cal P}\frac{1}{E-H_0} - i \pi \delta(E-H_0) \ ,
\label{eq:A3a}
\end{equation}
with ${\cal P}$ standing for the principal value part, while the propagator involving unstable particles is of the form \cite{MSL07,DORING2009170} ($\Pi =$ finite) 
\begin{align}
G & = \frac{1}{E - h_0 - \Pi} \nonumber \\
&= \frac{E-H_0}{(E-H_0)^2 + \Pi^{I\, 2}}  - i\frac{\Pi^I}{(E-H_0)^2 + \Pi^{I\, 2}} \ ,
\label{eq:A3b}
\end{align}
where $h_0$ denotes the unperturbed Hamiltonian involving the bare unstable particle and   
$\Pi$ is the self-energy of that unstable particle. $H_0 \equiv h_0 + \Pi^R$, with $\Pi = \Pi^R - i \Pi^I$.

Inserting Eq.~(\ref{eq:A2}) into Eq.~(\ref{eq:A1}), we have \cite{BKPS82}
\begin{align}
T  = K - i TG^IK  = K -iKG^IT \ , 
\label{eq:A4}
\end{align}
with the $K$-matrix ($K$) given by
\begin{equation}
K = V + VG^RK  = V + KG^RV 
\label{eq:A5}
\end{equation}
which is Hermitian if the driving potential $V$ is Hermitian.
  
For stable particles, using Eq.~(\ref{eq:A3a}), Eq.~(\ref{eq:A4}) becomes
\begin{equation}
T(E)  = K(E) - i\pi T(E) \delta(E-H_0) K(E) \ ,  
\label{eq:A6}
\end{equation}
which is the familiar equation for the $T$-matrix in terms of the $K$-matrix. Note that, for unstable particles propagation [cf. Eq,~(\ref{eq:A3b})], the imaginary part of $G$ -- for which there is no $\delta$-function in energy -- leads to a
momentum loop integration over the intermediate state.

Equation (\ref{eq:A1}) -- and consequently all the subsequent equations -- represents actually coupled equations in two-particle channels. Explicitly, for Eqs.~(\ref{eq:A4}) and (\ref{eq:A5}), we have (using the corresponding first equalities)
\begin{align}
T_{\alpha'\alpha} & = K_{\alpha'\alpha} - i \sum_\beta T_{\alpha'\beta} G^I_\beta K_{\beta\alpha} \ , \nonumber \\
K_{\alpha'\alpha} & = V_{\alpha'\alpha} +  \sum_\beta K_{\alpha'\beta} G^R_\beta V_{\beta\alpha} \ , 
\label{eq:A7}
\end{align}
where the subscripts stand for the two-particle channels, i.e., $\alpha'$ denotes the final two-particle channel and $\alpha$, the initial two-particle channel. $\beta$ denotes the intermediate two-particle channel and it is summed over all the channels (including the stable- and unstable-particles propagations) to account for the possible couplings of the initial and final states to all other channels. 
Note that, as explained in Sec.~\ref{sec:Notation},  the equations in (\ref{eq:A7}) may be interpreted as given either in plane-wave or in partial-wave basis. For the latter, the indices $\alpha'$, $\alpha$ and $\beta$ specify also the partial-wave states, in addition to the two-particle channels.  Note also that the reference to the two-particle relative momentum is completely suppressed in the present notation, including the momentum-loop integration over the intermediate states.

Usually, the integral equation for $T$ in Eq.~(\ref{eq:A7}) is solved to yield
\begin{equation}
T_{\alpha'\alpha} = \sum_{\beta }K_{\alpha'\beta} \left[\frac{1}{1+i G^I K}\right]_{\beta\alpha}   \ .
\label{eq:A4a}
\end{equation}

In the present work, however, we solve that equation as follows. First, we write it as
\begin{align}
\sum_{\beta \ne \alpha'} T_{\alpha'\beta} ( \delta_{\beta\alpha} + i G^I_\beta K_{\beta\alpha}) & = (1 - i T_{\alpha'\alpha'} G^I_{\alpha'} ) K_{\alpha'\alpha} \nonumber \\
\sum_{\beta \ne \alpha'} T_{\alpha'\beta} D_{\beta\alpha} & = N_{\alpha'} K_{\alpha'\alpha} \ , 
\label{eq:A8x}
\end{align}
where we have defined
\begin{align}
D_{\beta'\beta} & \equiv \delta_{\beta'\beta} + iG^I_{\beta'} K_{\beta'\beta} \ , \ \ \ \ \ (\beta',\beta \ne \alpha') \nonumber \\
N_{\alpha'} & \equiv 1 - iT_{\alpha'\alpha'}G^I_{\alpha'} \ .
\label{eq:A9}
\end{align}
Next, we multiply Eq.~(\ref{eq:A8x}) throughout from the right by the inverse matrix of $D$ to get
\begin{equation}
T_{\alpha'\alpha} = N_{\alpha'} \sum_{\beta' \ne \alpha'} K_{\alpha'\beta'} D^{-1}_{\beta'\alpha} \ .
\label{eq:A8y}
\end{equation}
Finally, we insert the above result back into the equation for $T$ in (\ref{eq:A7}) to arrive at
\begin{align}
T_{\alpha'\alpha} & = N_{\alpha'}\Big[ K_{\alpha'\alpha} - i\sum_{\beta,\beta'\ne\alpha'} K_{\alpha'\beta'}\left(D^{-1}\right)_{\beta'\beta}G^I_\beta K_{\beta\alpha}\Big] \nonumber \\
& = N_{\alpha'}\hat{K}_{\alpha'\alpha} \ .
\label{eq:A8}
\end{align}

The last equality in Eq.~(\ref{eq:A8}) defines $\hat{K}$ to be 
\begin{equation}
\hat{K}_{\alpha'\alpha} \equiv K_{\alpha'\alpha}  - i\sum_{\beta,\beta'\ne\alpha'} K_{\alpha'\beta'} \left(D^{-1}\right)_{\beta'\beta} G^I_\beta K_{\beta\alpha} \ ,
\label{eq:Khat}
\end{equation}
which -- unlike the $K$-matrix -- is, in general, a complex quantity.  Note that below the first inelastic threshold, $\hat{K} = K$.
Also, note that the explicit dependence on the channel $\alpha'$ in the intermediate state is absent in $\hat{K}_{\alpha'\alpha}$. This dependence is contained implicitly in the $K$-matrices, $K_{ij}$. 

Inserting Eq.~(\ref{eq:A8}) into  the definition of $N_{\alpha'}$ in Eq.~(\ref{eq:A9}), yields
\begin{equation}
N_{\alpha'} = 1 - iT_{\alpha'\alpha'}G^I_{\alpha'} = \frac{1}{1 + i\hat{K}_{\alpha'\alpha'} G^I_{\alpha'} }\ .
\label{eq:A9a}
\end{equation}

\vskip 0.5cm
Starting from the second equality in Eq.~(\ref{eq:A4}), it is straightforward to show that the $T$-matrix can be also expressed as
\begin{equation}
T_{\alpha'\alpha} = \hat{\bar{K}}_{\alpha'\alpha} \bar{N}_{\alpha}  \ ,
\label{eq:A13}
\end{equation}
where
\begin{align}
\bar{N}_{\alpha} & \equiv 1 - i G^I_\alpha T_{\alpha\alpha} = \frac{1}{1 + i G^I_{\alpha} \hat{K}_{\alpha\alpha}  } \ , \nonumber \\
\bar D_{\beta'\beta} & \equiv \delta_{\beta'\beta} + i K_{\beta'\beta} G^I_{\beta}  \ , \ \ \ \ \ (\beta',\beta \ne \alpha) \nonumber \\
\hat{\bar{K}}_{\alpha'\alpha} & \equiv K_{\alpha'\alpha}  - i\sum_{\beta,\beta'\ne\alpha} K_{\alpha'\beta'} G^I_{\beta'} \left(\bar D^{-1}\right)_{\beta'\beta} K_{\beta\alpha} \ .
\label{eq:14}
\end{align}

Equation (\ref{eq:A8}) or (\ref{eq:A13}) is the desired result: we have exhibited the full phase structure of the $T$-matrix which is nontrivial in general due to the phase structure of $\hat{K}_{\alpha'\alpha}$, introduced by the terms involving $G^I_\beta$s in Eq.~(\ref{eq:Khat}) or (\ref{eq:14}). 
For on-shell $\hat{K}_{\alpha\alpha}$, its phase structure can be expressed simply in terms of the phase-shift and inelasticity of the elastic scattering $T$-matrix as shown in Appendix \ref{ap:PSparam}.

Note also that Eq.~(\ref{eq:A8}) or (\ref{eq:A13}) is completely general and holds for fully off-shell $T$-matrix.
Hereafter, we refer to the factors $N_{\alpha'}$ and $\bar{N}_\alpha$ defined in Eqs.~(\ref{eq:A9}) and (\ref{eq:14}) as the \textit{generalized} Watson's factors. For completeness, we show how Watson's theorem emerges from these equations in the following Sec.~\ref{subset:PS1-TC}, when the initial channel $\alpha$ corresponds to the photon-baryon channel.

If we wish, combining Eqs.~(\ref{eq:A8}) and (\ref{eq:A13}), the $T$-matrix can be expressed in a symmetric form
\begin{equation}
T_{\alpha'\alpha} = \frac12 \left( N_{\alpha'}\hat{K}_{\alpha'\alpha}  + \hat{\bar{K}}_{\alpha'\alpha} \bar{N}_{\alpha} \right)  \ .
\label{eq:A8-13}
\end{equation}

\subsection{Two-channel case and Watson's theorem}\label{subset:PS1-TC}

Confining now to the case of two channels problem, $\hat{K}_{\alpha'\alpha}$ in Eq.~(\ref{eq:Khat}) simplifies and Eq.~(\ref{eq:A8}) takes the form
\begin{equation}
T_{\alpha'\alpha} = N_{\alpha'} \left[ K_{\alpha'\alpha} - iK_{\alpha'\beta} \bar{N}_{K\beta} G^I_\beta K_{\beta\alpha} \right] \ , \ \ \ (\beta \ne \alpha') 
\label{eq:A15}
\end{equation}
with
\begin{equation}
\bar{N}_{K\alpha}  \equiv  \frac{1}{1 + i G^I_\alpha K_{\alpha\alpha}} \ .
\label{eq:A17a}
\end{equation}

For a transition reaction, where $\alpha'\ne\alpha$, Eq.~(\ref{eq:A15}) further reduces to
\begin{equation}
T_{\alpha'\alpha}  = N_{\alpha'} K_{\alpha'\alpha} \bar{N}_{K\alpha} \ .
\label{eq:A17}
\end{equation}

If the two channels considered involve only stable particles, then, in partial-wave basis, Eqs.~(\ref{eq:A15}) and (\ref{eq:A17}) are simple algebraic equations, where $G^I_{\beta} \to \rho_{\beta}$ after the momentum loop integration with $\rho_\beta$ denoting the phase-space density.
Moreover, if the on-shell $T$-matrix and the on-shell $K$-matrix can be parametrized in terms of phase-shifts and inelasticities as given in Appendix \ref{ap:PSparam}, we obtain from Eq.~(\ref{eq:A17}), 
\begin{equation}
T_{\alpha' \alpha}  = \left\{\frac12 \left(\eta_{\alpha'} e^{i2\delta_{\alpha'}} + 1\right)\right\} K_{\alpha'\alpha} \left(e^{i\delta_\alpha}\cos\delta_\alpha \right) \,,  
\label{eq:A15b}
\end{equation}
for the transition amplitude ($\alpha' \ne \alpha$). Here, we have made use of Eqs.~(\ref{eq:D3}) and (\ref{eq:D3a}).

Equation (\ref{eq:A15b}) reveals the phase structure of the $T$-matrix amplitude explicitly in terms of the phase-shifts for the transition amplitude in the case of a two-channel problem.
It is the analog of the well-known Watson's theorem for photoproduction \cite{Watson54} in the case of two-body hadronic reactions.  The phase of the reaction amplitude is determined by both the on-shell initial and final state interactions through the Watson's factors $\bar{N}_{K\alpha}$ and $N_{\alpha'}$, respectively.
Note that, in Eq.~(\ref{eq:A15b}), the effect of the channel openings is lumped entirely into the final state interaction factor.
We remind that, from Eq.~(\ref{eq:A5}), if $V$ is Hermitian, so is $K$ and, together with time reversal invariance, $K_{\alpha'\alpha}$ is either pure real or pure imaginary.

If we start with the $T$-matrix in the form given by Eq.~(\ref{eq:A13}), instead of Eq.~(\ref{eq:A8}) as we have done above, we obtain an equivalent alternative form for the transition  amplitude ($\alpha' \ne \alpha$),
\begin{equation}
T_{\alpha'\alpha}  = N_{\alpha'} K_{\alpha'\alpha} \bar{N}_{\alpha} \ ,
\label{eq:A19}
\end{equation}
with
\begin{equation}
N_{K\alpha}  \equiv  \frac{1}{1 + i K_{\alpha\alpha} G^I_\alpha } \ .
\label{eq:A19a}
\end{equation}

In terms of the phase-shift parametrization, Eq.~(\ref{eq:A19}) becomes
\begin{equation}
T_{\alpha' \alpha}  = \left(e^{i\delta_{\alpha'}}\cos\delta_{\alpha'} \right) K_{\alpha'\alpha} \left\{\frac12 \left(\eta_{\alpha} e^{i2\delta_{\alpha}} + 1\right)\right\}  \ . 
\label{eq:A19b}
\end{equation}
In contrast to Eq.~(\ref{eq:A15b}), where the effect of the channel openings is lumped into the final state interaction factor, here, this effect is lumped into the initial state interaction factor.

It should be mentioned that, strictly speaking,  the two channels consideration of the meson-baryon reaction processes applies only to $\pi N$ charge-exchange scatterings, such as $\pi^0 p \to \pi^+ n$. This is due to the fact that the lightest  meson-baryon channel, apart from $\pi N$, is the $\eta N$ channel which is already above the $\pi\pi N$ threshold, leading to the presence of an inelastic channel even when the isospin symmetry breaking of the strong interaction is ignored.

\vskip 0.5cm
In the case of meson photoproduction, Eq.~(\ref{eq:A17}) becomes ($\alpha' \ne \alpha=\gamma$)
\begin{equation}
T_{\alpha'\gamma} =   N_{\alpha'} K_{\alpha'\gamma}\, \bar{N}_{K\gamma} \ , 
\label{eq:A18}
\end{equation}
where 
$\bar{N}_{K\gamma} = 1/(1 + i G^I_\gamma K_{\gamma\gamma})$ is the Watson's factor due to the $\gamma N$ initial state interaction. 
In the one-photon approximation, due to the weakness of the electromagnetic interaction,  the Watson's factor $N_{K\gamma}$ approaches unit since we may set $K_{\gamma\gamma}$  appearing in $N_{K\gamma}$ to zero. Likewise, for the two-channel case, where one of the channels is the photon-baryon channel, $N_{\alpha'}=N_{K\alpha'}$ in one-photon approximation. 
Equation (\ref{eq:A18}), then, becomes
\begin{equation}
T_{\alpha'\gamma} =  N_{K\alpha'} K_{\alpha'\gamma} = \left(e^{i\delta_{\alpha'}} \cos\delta_{\alpha'} \right) K_{\alpha'\gamma} \ , 
\label{eq:A18a}
\end{equation}
which is the usual form of Watson's theorem for photoproduction \cite{Watson54}. Equation (\ref{eq:A19}) yields the same result as above.
Note that Watson's theorem is a direct consequence of unitarity and time reversal invariance of the $S$-matrix, in addition to the one-photon approximation assumption.  Also, as is well known, in practice, ignoring the isospin symmetry breaking of the hadronic interactions, Watson's theorem applies to pion photoproduction below $\pi\pi N$ threshold.

\section{Phase structure of the pole and nonpole meson-baryon $T$-matrix}\label{sec:PS2}

In this section we exhibit the phase structure of the $T$-matrix in terms of the pole ($T^P$) and nonpole ($X\equiv T^{NP}$) parts.

First, we recall that the full $T$-matrix given by Eq.~(\ref{eq:A1}) can be decomposed as (see, Appendix~\ref{ap:C})
\begin{align}
T & = V + VGT \nonumber \\
& \equiv T^P + X \ ,
\label{eq:F1}
\end{align}
where
\begin{equation}
X = U + UGX \ ,
\label{eq:F2}
\end{equation}
with $U\equiv V^{NP} \equiv V-V^P$ and
\begin{equation}
V^P = \sum_r \ket{F_{0r}} S_{0r} \bra{F_{0r}} \ ,
\label{eq:F2a}
\end{equation}
where $\ket{F_{0r}}$ and $S^{-1}_{0r} = E - m_{0r}$ denote the bare meson-baryon vertex and bare baryon propagator, respectively. The summation is over the resonances specified by index $r$.

 The pole part of the $T$-matrix in Eq.~(\ref{eq:F1}) is (following the ket and bra notation used in Ref.~\cite{H97,HNK06})
\begin{equation}
T^P = \sum_{r'r} \ket{F_{r'}} S_{r'r} \bra{F_r} \ , 
\label{eq:F3}
\end{equation}
where the dressed vertices  read 
\begin{align}
\ket{F}_{r'} & \equiv \left(1 + XG\right) \ket{F_{0r'}} \ , \nonumber \\
\bra{F}_r & \equiv \bra{F_{0r}} \left(1 + GX \right) \ ,   
\label{eq:F4}
\end{align}
and the dressed baryon propagator, $S_{r'r}$,
\begin{equation}
S_{r'r}^{-1} = S_{0r}^{-1} \delta_{r'r}  - \Sigma_{r'r}
\label{eq:F5}
\end{equation}
with the self-energy
\begin{equation}
\Sigma_{r'r} \equiv  \bra{F_{0r'}} G \ket{F_r}  \ .
\label{eq:F6}
\end{equation}
Note that the dressed baryon propagator in Eq.~(\ref{eq:F5}) couples resonances, so it is a matrix propagator in resonance space.  It's structure is shown explicitly in Appendix \ref{ap:TRC} for the case of a two-resonance coupling since, in practice, this is the maximum number of resonance couplings in most of the cases. 

Second, since the structures of the $T$- and $K$-matrix scattering equations are the same [cf.~Eqs.~(\ref{eq:A1},\ref{eq:A5})], it is straightforward to decompose 
the $K$-matrix into the pole ($K^P$) and nonpole ($W\equiv K^{NP}$) parts
\begin{align}
K & = V + VG^RK \nonumber \\
& \equiv K^P + W  \ ,
\label{eq:F7}
\end{align}
where
\begin{equation}
W  = U  + U G^R W \ ,
\label{eq:F8}
\end{equation}
and
\begin{equation}
K^P  = \sum_{r'r} \ket{F_{Kr'}} S_{Kr'r} \bra{F_{Kr}} \ .
\label{eq:F9}
\end{equation}
Here, the dressed vertices are given by
\begin{align}
\ket{F_{Kr'}} & \equiv \left(1  + W G^R \right)\ket{F_{0r'}} \ , \nonumber \\ 
\bra{F_{Kr}} & \equiv \bra{F_{0r}} \left(1 + G^R W \right) \ ,
\label{eq:F10}
\end{align}
and the dressed baryon propagator by
\begin{equation}
S_{Kr'r}^{-1} = S_{0r}^{-1}\delta_{r'r} - \Sigma_{Kr'r} \ ,
\label{eq:F11}
\end{equation}
with the self-energy
\begin{equation}
\Sigma_{K r'r} = \bra{F_{0r'}} G^R  \ket{F_{Kr}} \ .
\label{eq:F12}
\end{equation}

Third, since the $T$-matrix can be expressed in terms of the $K$-matrix  as given by Eq.~(\ref{eq:A4}), which exhibits the same integral-equation structure as Eq.~(\ref{eq:A1}), except for the appearance of the imaginary part of the meson-baryon propagator $-iG^I$ instead of the full propagator $G$, it is straightforward to express the pole and nonpole $T$-matrices [cf. Eqs.(\ref{eq:F1}), (\ref{eq:F2}), (\ref{eq:F3})-%,\ref{eq:F4},\ref{eq:F5},
(\ref{eq:F6})] in terms of the $K$-matrix [cf. Eqs.(\ref{eq:F7})-%,\ref{eq:F8},\ref{eq:F9},\ref{eq:F10},\ref{eq:F11},
(\ref{eq:F12})]. Then, the nonpole $T$-matrix $X (\equiv T^{NP})$ given by Eq.~(\ref{eq:F2}) becomes
\begin{equation}
X  = W - i XG^I W  = W -iWG^IX \ .
 \label{eq:F13}
 \end{equation}

The pole part ($T^P$) is given by Eq.~(\ref{eq:F3}) with
 \begin{align}
 \ket{F_{r'}} & \equiv (1 -i  XG^I )\ket{F_{Kr'}}  \ , \nonumber \\
 \bra{F_r} & \equiv \bra{F_{Kr}} (1 - i G^IX )  \ ,
\label{eq:F14}
\end{align}
and the dressed propagator $S_{r'r}$ expressed as 
\begin{equation}
S_{r'r}^{-1} = S_{Kr'r}^{-1} - \hat{\Sigma}_{r'r}
\label{eq:F27}
\end{equation}
where the self-energy $\hat{\Sigma}$ is
\begin{equation}
\hat{ \Sigma}_{r'r} \equiv -i \bra{F_{Kr'}} G^I \ket{F_r}  \ .
 \label{eq:F15}
 \end{equation}

\vskip 0.5cm
Writing the meson-baryon channel indices explicitly, we have, for Eq.~(\ref{eq:F8}),
\begin{equation}
W_{\alpha'\alpha}  = U_{\alpha'\alpha} +  \sum_\beta W_{\alpha'\beta} G^R_\beta U_{\beta\alpha} \ . 
\label{eq:F8a}
\end{equation}

For Eq.~(\ref{eq:F13}), we have,
\begin{align}
X_{\alpha'\alpha} & = W_{\alpha'\alpha} - i \sum_\beta X_{\alpha'\beta} G^I_\beta W_{\beta\alpha}  \nonumber \\
                            & = W_{\alpha'\alpha} - i \sum_\beta W_{\alpha'\beta} G^I_\beta X_{\beta\alpha}  \ ,
\label{eq:F13a}
\end{align}
which can be solved to yield (from the first equality)
\begin{equation}
X_{\alpha'\alpha}  = N^X_{\alpha'} \hat{W}_{\alpha'\alpha} \ , 
\label{eq:F16}
\end{equation}
with
\begin{equation}
\hat{W}_{\alpha'\alpha} \equiv W_{\alpha'\alpha}  - i\sum_{\beta,\beta'\ne\alpha'} W_{\alpha'\beta'} \left({(D^X)}^{-1}\right)_{\beta'\beta} G^I_\beta W_{\beta\alpha} \ ,
\label{eq:F17}
\end{equation}
and
\begin{align}
D^X_{\beta'\beta} & \equiv \delta_{\beta'\beta} + iG^I_{\beta'} W_{\beta'\beta} \ , \nonumber \\
N^X_{\alpha'} & \equiv 1 - iX_{\alpha'\alpha'}G^I_{\alpha'} = \frac{1}{1 + i\hat{W}_{\alpha'\alpha'}G^I_{\alpha'} } \ .
\label{eq:F18}
\end{align}
%

%\cor{$\hat{W}_{\alpha'\alpha}$ can be \cor{explicitly decomposed into it's real and imaginary parts and Eq.~(\ref{eq:F17}) can be rewritten as}{decomposes as}}{In the particular case where $W_{\alpha'\alpha}$ are pure real for all the channels $\alpha', \alpha$ considered in the problem, $\hat{W}_{\alpha'\alpha}$ in Eq.~(\ref{eq:F17}) can be decomposed into it's real and imaginary parts as}
%%
%\begin{equation}
%\hat{W}_{\alpha'\alpha} = \hat{W}^R_{\alpha'\alpha} + i \hat{W}^I_{\alpha'\alpha}  \ ,
%\label{eq:F17a}
%\end{equation}
%%
%where
%%
%\begin{align}
%\hat{W}^R_{\alpha'\alpha} & \equiv W_{\alpha'\alpha}  \nonumber \\
%&  - \sum_{\beta,\beta'\ne\alpha'} W_{\alpha'\beta'} \left(G^IW \frac{1}{1 + (G^IW)^2}\right)_{\beta'\beta} G^I_\beta W_{\beta\alpha} \ , \nonumber \\
%\hat{W}^I_{\alpha'\alpha} & \equiv - \sum_{\beta,\beta'\ne\alpha'} W_{\alpha'\beta'} \left(\frac{1}{1 + (G^IW)^2}\right)_{\beta'\beta} G^I_\beta W_{\beta\alpha} \  .
%\nonumber \\
%\label{eq:F17b}
%\end{align}
%%
%\cor{\cor{In the above equations, $|G^IW|^2 = (G^I W)(G^I W)^\dagger$.}{In the above equations, $\hat{W}^{\cal A}$ ($\hat{W}^{\cal B}$) is either pure real when $W_{\alpha'\alpha}$ is pure real or pure imaginary when $W_{\alpha'\alpha}$ is pure imaginary.}}{}.

\vskip 0.5cm
From the second equality in Eq.~(\ref{eq:F13a}), it is also immediate that $X_{\alpha'\alpha}$ can be expressed as
\begin{equation}
X_{\alpha'\alpha}  =  \hat{\bar{W}}_{\alpha'\alpha} \bar{N}^X_{\alpha}  \ ,
\label{eq:F20}
\end{equation}
with
\begin{equation}
\hat{\bar{W}}_{\alpha'\alpha} \equiv W_{\alpha'\alpha}  - i\sum_{\beta,\beta'\ne\alpha} W_{\alpha'\beta'} G^I_{\beta'} \left({(\bar{D}^X)}^{-1}\right)_{\beta'\beta}  W_{\beta\alpha}  \ ,
\label{eq:F20a}
\end{equation}
and
\begin{align}
\bar{D}^X_{\beta'\beta} & \equiv \delta_{\beta'\beta} + i W_{\beta'\beta}G^I_\beta \ , \nonumber \\
\bar{N}^X_{\alpha} & \equiv 1 - i G^I_\alpha X_{\alpha\alpha} = \frac{1}{1 + iG^I_\alpha\hat{\bar{W}}_{\alpha\alpha}} \ .
\label{eq:F21}
\end{align}
Note that in the case the particles in channel $\alpha$ are stable ($G^I_\alpha \to \rho_\alpha$), $\bar{N}^X_{\alpha} = N^X_{\alpha}$.

%Equation (\ref{eq:F20a}) can also be rewritten as
%%
%\begin{equation}
%\hat{\bar{W}}_{\alpha'\alpha} = \hat{\bar{W}}^R_{\alpha'\alpha} + i \hat{\bar{W}}^I_{\alpha'\alpha}  \ ,
%\label{eq:F20b}
%\end{equation}
%%
%\cor{where}{in the case $W_{\alpha'\alpha}$ are pure real for all the channels $\alpha', \alpha$ considered, with}
%%
%\begin{align}
%\hat{\bar{W}}^R_{\alpha'\alpha} & \equiv W_{\alpha'\alpha}  \nonumber \\
%&  - \sum_{\beta,\beta'\ne\alpha} W_{\alpha'\beta'} G^I_{\beta'}\left( \frac{1}{1 + (WG^I)^2} WG^I\right)_{\beta'\beta} W_{\beta\alpha} \ , \nonumber \\
%\hat{\bar{W}}^I_{\alpha'\alpha} & \equiv - \sum_{\beta,\beta'\ne\alpha} W_{\alpha'\beta'} G^I_{\beta'}\left(\frac{1}{1 + (WG^I)^2}\right)_{\beta'\beta} W_{\beta\alpha} \ .
%\nonumber \\
%\label{eq:F20c}
%\end{align}
%%

In the following, to exhibit the phase structure of the pole $T$-matrix, $T^P$, we make use of the dressed vertices and propagator as given by Eqs.~(\ref{eq:F14},\ref{eq:F27},\ref{eq:F15}). Writing the meson-baryon channel indices explicitly, Eq.~(\ref{eq:F3}) becomes
\begin{equation}
T^P_{\alpha'\alpha} = \sum_{r'r} \ket{F_{r'}}_{\alpha'} S_{r'r} \bra{F_r}_\alpha \ .
\label{eq:F3a}
\end{equation}
The dressed meson-baryon vertex $\ket{F_{r'}}$ [cf. Eq.~(\ref{eq:F14})] becomes
\begin{align}
\ket{F_{r'}}_{\alpha'} & \equiv \ket{F_{Kr'}}_{\alpha'}  - i \sum_{\beta} X_{\alpha'\beta}G^I_{\beta}\ket{F_{Kr'}}_\beta \nonumber \\
& =  N^X_{\alpha'} \ket{\hat{F}_{Kr'}}_{\alpha'} \ ,
\label{eq:F14a}
\end{align}
where
\begin{equation}
\ket{\hat{F}_{Kr'}}_{\alpha'} \equiv \ket{F_{Kr'}}_{\alpha'}  - i \sum_{\beta\ne\alpha'}\hat{W}_{\alpha'\beta}G^I_{\beta}\ket{F_{Kr'}}_\beta \ .
\label{eq:F22}
\end{equation}
To arrive at the last equality in Eq.~(\ref{eq:F14a}), Eqs.~(\ref{eq:F18}) and (\ref{eq:F16}) have been used.

Analogously,
\begin{align} 
\bra{F_r}_\alpha & \equiv  \bra{\hat{F}_{Kr}}_\alpha \bar{N}^X_{\alpha}  \ ,
\label{eq:F24}
\end{align}
where
\begin{equation}
\bra{\hat{F}_{Kr}}_\alpha \equiv \bra{F_{Kr}}_\alpha  - i \sum_{\beta\ne\alpha} \bra{F_{Kr}}_\beta G^I_{\beta} \hat{\bar{W}}_{\beta\alpha} \ .
\label{eq:F25}
\end{equation}

The self-energy given by Eq.~(\ref{eq:F15}) reads
\begin{align}
 \hat{\Sigma}_{r'r} & = -i \sum_\beta \bra{F_{Kr'}}_\beta G^I_\beta \ket{F_r}_\beta \nonumber \\
 & =  -i \sum_\beta \bra{F_{Kr'}}_\beta G^I_\beta N^X_{\beta} \ket{\hat{F}_{Kr}}_\beta \ .
 \label{eq:F15a}
 \end{align}

Then, inserting the above result into Eq.~(\ref{eq:F27}), we have for the full propagator,
\begin{equation}
S^{-1}_{r'r}  =  S^{-1}_{K r'r}  + i \sum_\beta \bra{F_{Kr'}}_\beta G^I_\beta N^X_{\beta} \ket{\hat{F}_{Kr}}_\beta \ .
\label{eq:F30}
\end{equation}

Finally, making use of Eq.~(\ref{eq:F11}) and inserting Eqs.~(\ref{eq:F14a}), (\ref{eq:F24}), and (\ref{eq:F30}) into Eq.~(\ref{eq:F3a}), and combining with Eq.~(\ref{eq:F16}), we arrive at the result we are seeking
\begin{widetext}
\begin{align}
T_{\alpha'\alpha} & = T^P_{\alpha'\alpha} + X_{\alpha'\alpha} \nonumber \\
& = \sum_{r'r} \left\{ N^X_{\alpha'}\ket{\hat{F}_{Kr'}}_{\alpha'} \left(\frac{1}{(E - m_{0})I - \Sigma_K + i \sum_\beta \bra{F_K}_{\beta} G^I_\beta N^X_{\beta}\ket{\hat{F}_K}_{\beta}   }\right)_{r'r} \bra{\hat{F}_{Kr}}_\alpha \bar{N}^X_{\alpha} \right\}  + N^X_{\alpha'} \hat{W}_{\alpha'\alpha} \ .
\label{eq:F28}
\end{align}
\end{widetext}
where $I$ stands for the identity matrix in resonance space.  $\Sigma_K$ is given by Eq.~(\ref{eq:F12}).                       
                         
The above equation exhibits the full phase structure of the $T$-matrix amplitude in terms of the pole and nonpole parts. First of all, we note that the phase structure of the $T$-matrix is determined by the branch points introduced in the amplitude due to the opening of the meson-baryon channels. This is controlled by the availability of the phase space for a given meson-baryon channel $\beta$ encoded in the imaginary part of the corresponding meson-baryon propagator $G^I_\beta$.  This quantity appears implicitly in many places in Eq.~(\ref{eq:F28}) and, consequently, makes the phase structure of the $T$-matrix highly nontrivial in general. Note that the Watson's factor $N^X$ and all the quantities with ``hat" in Eq.~(\ref{eq:F28}) involve $G^I$ [cf. Eqs.~(\ref{eq:F16}), (\ref{eq:F18}), (\ref{eq:F20a}), (\ref{eq:F21}), (\ref{eq:F22}), and (\ref{eq:F25})]. All other terms appearing in Eq.~(\ref{eq:F28}) are real quantities and do not involve $G^I$.  We also recall that the dressed vertex $\ket{\hat{F}_K}$($\bra{\hat{F}_K}$),  as well as the Watson's factor $N^X$($\bar{N}^X$), are all expressed in terms of the quantity $\hat{W}$($\hat{\bar{W}}$) [cf. Eqs.~(\ref{eq:F18}), (\ref{eq:F21}), (\ref{eq:F22}), and (\ref{eq:F25})]. The latter quantity is the nonpole $T$-matrix apart from the Watson's factor $N^X$ [cf. Eqs.~(\ref{eq:F16}) and (\ref{eq:F20})]. This means that the dynamical effects on the phase structure are determine by the nonpole part of the $T$-matrix (up to the corresponding Watson's factor), and that there is an intimate relationship between the phase structure of the pole and nonpole parts of the $T$-matrix amplitude.

\vskip 0.3cm
In the following, we discuss the elastic scattering below the first inelastic threshold where the phase structure of the $T$-matrix amplitude becomes much simpler. 
Here, we ignore the resonance couplings for simplicity.  We also assume a stable meson-baryon channel $\alpha$ and consider the phase-shift parametrization of the on-shell nonpole $T$-matrix such that $N^X_\alpha = \bar{N}^X_\alpha = e^{\delta^X_\alpha}\cos\delta^X_\alpha$, where $\delta^X_\alpha$ stands for the phase-shift of the nonpole $T$-matrix ($X \equiv T^{NP} $). Then, in partial-wave basis, Eq.~(\ref{eq:F28}) reduces to
\begin{equation}
T_{\alpha\alpha} =  \sum_{r} \left\{ e^{i\delta^X_{\alpha}}  g_{\alpha r}
\frac{1}{E - M_r  + i \frac{\Gamma_r}{2}} 
g_{\alpha r} e^{i\delta^X_\alpha} \right\}  + e^{i\delta^X_{\alpha}} \tilde{W}_{\alpha\alpha} \ ,
\label{eq:E2}
\end{equation}
where we have introduced the (suggestive) notations
\begin{align}
g_{\alpha r} & \equiv \cos\delta^X_{\alpha}\, \ket{F_K}_{\alpha, r} \ , \nonumber \\
\Gamma_r & \equiv  2 \rho_{\alpha}\,  g^2_{\alpha r}  \ , \nonumber \\
M_r & \equiv  m_{0r}  + \Sigma_{Krr}  + \tan\delta^X_{\alpha}\, \frac{\Gamma_r}{2} \ , \nonumber \\
\tilde{W}_{\alpha\alpha} & \equiv \cos\delta^X_{\alpha} W_{\alpha\alpha} \ . 
\label{eq:E3}
\end{align}
Equation (\ref{eq:E2}) exhibits, explicitly, the full phase structure of the elastic $T$-matrix amplitude below the first inelastic threshold. Apart from the phase $e^{i\delta^X_\alpha}$ arising from the Watson's factors in the dressed vertices and propagator, there is also the same phase factor arising from the Watson's factor in the nonpole part of the amplitude. Note that the last term in Eq.~(\ref{eq:E2}) is simply the statement of Watson's theorem for the nonpole $T$-matrix $X$. Recall that $W$ is the nonpole part of the $K$-matrix and, as such, it is Hermitian if the nonpole driving potential $U \equiv V^{NP} $ is.

%For completeness, the reaction amplitude (\ref{eq:F28}) for the case of two coupled channels is given in Appendix~\ref{ap:PS2-TC}. 

\vskip 0.3cm
Equation (\ref{eq:F28}) is the main result of this section. It
serves as a convenient starting point for approximations one can make with varying degrees of sophistication. In particular, it allows to keep track on the basic properties of the $S$-matrix in these approximations. Indeed, Eq.~(\ref{eq:F28}) is being used by us in the construction of an isobar model in which unitarity is automatically satisfied.

\section{Phase Structure of the Photoproduction amplitude}\label{sec:PS3}

As shown explicitly in Appendix~\ref{ap:C}, the gauge-invariant photoproduction amplitude in one-photon approximation also admits a decomposition  into the pole and nonpole parts. Thus, we must be able to exhibit the complex phase structure of this amplitude in terms of the corresponding pole and nonpole amplitudes, analogous to what has been done for the meson-baryon $T$-matrix amplitude in the previous section.
\footnote{Note that, to preserve gauge invariance of the decomposed photoproducton amplitude into the pole and nonpole parts, we need to consider what to take for the nonpole driving potential $U^\mu  (\equiv V^{NP\, \mu})$ and for the bare photon coupling $\bra{F^\mu_{0r}}$. They enter in the definition of $W^\mu$ and $\bra{F^\mu_{Kr}}$ in Eqs.~(\ref{eq:PF11}) and (\ref{eq:PF13}), respectively.
For example, in the field theoretic approach of Appendix~\ref{ap:C},  the bare coupling $\bra{F^\mu_{0r}}$ gets renormalized as given by Eq.~(\ref{eq:H4}). And the driving potential $U^\mu$  contains additional terms compared to the usual $u$- and $t$-channel Feynman diagrams [cf.~Eq.~(\ref{eq:H7})]. These observations should be kept in mind when constructing (gauge-invariant) photoproduction amplitude in the present approach.}
Indeed, the meson photoproduction amplitude can be obtained by simply considering the photon-baryon channel as an additional channel in the coupled channels $T$-matrix equation of Eq.~(\ref{eq:F1}), i.e., all the results of the previous sections apply to photoproduction as well. 
%In practice, however, one usually calculates the photoproduction amplitude in one-photon approximation due to the weakness of the electromagnetic interaction. 
In terms of the coupled channels formulation of the previous sections, the one-photon approximation
means to ignore the photon-baryon channel in the intermediate states, i.e, this channel appears only as the initial state. Then, Eq.~(\ref{eq:F28}) leads to
\begin{widetext}
\begin{align}
M^\mu_{\alpha'\gamma} & = M^{P\, \mu}_{\alpha'\gamma} + X^\mu_{\alpha'\gamma} \nonumber \\
& = \sum_{r'r} \left\{ N^X_{\alpha'}\ket{\hat{F}_{Kr'}}_{\alpha'}  \left(\frac{1}{(E - m_{0})I - \Sigma_K + i \sum_\beta \bra{F_K}_{\beta} G^I_\beta N^X_{\beta}\ket{\hat{F}_K}_{\beta}   }\right)_{r'r}
 \bra{\hat{F}^\mu_{Kr}}_\gamma \right\}  + N^X_{\alpha'} \hat{W}^\mu_{\alpha'\gamma} \ ,
\label{eq:PF28}
\end{align}
\end{widetext}
where the initial meson-baryon channel $\alpha$ has been replaced by the photon-baryon channel $\gamma$ which appears only in the initial state. 
In particular, note that the Watson's factor $\bar{N}^X_\gamma  = 1/(1+iG^I_\gamma \hat{\bar{W}}^\mu_{\gamma\gamma}) \to 1$ in one-photon approximation. The superscript $\mu$ stands for the Lorentz index of the photon polarization.

The quantity $\hat{W}^\mu_{\alpha'\gamma}$ in Eq.~(\ref{eq:PF28}) follows from Eq.~(\ref{eq:F17}). Explicitly, we have
\begin{equation}
\hat{W}^\mu_{\alpha'\gamma}  \equiv W^\mu_{\alpha'\gamma}  - i \sum_{\beta,\beta'\ne\alpha'} W_{\alpha'\beta'} \left(\left(D^X\right)^{-1}\right)_{\beta'\beta} G^I_\beta W^\mu_{\beta\gamma}  \ .
\label{eq:PF17}
\end{equation}
where, from Eq.~(\ref{eq:F8a}),
 \begin{equation}
 W^\mu_{\alpha'\gamma}  = U^\mu_{\alpha'\gamma} + \sum_\beta W_{\alpha'\beta} G^R_\beta U^\mu_{\beta\gamma}  \ .
  \label{eq:PF11}
 \end{equation}
Note that the summations over the channels in the above two equations, and all the subsequent equations in this section, exclude the photon-baryon channel in the intermediate states, i.e., $\beta,  \beta' \ne \gamma$ due to the one-photon approximation. This is to be understood  for the remainder of this paper.
 %
%Following  Eqs.~(\ref{eq:F17a},\ref{eq:F17b}), \cor{}{in the case $W_{\alpha'\alpha}$ and $W_{\beta\gamma}$  are pure real for all the channels considered,} Eq.~(\ref{eq:PF17}) can be re-expressed as 
% %
%\begin{equation}
%\hat{W}^\mu_{\alpha'\gamma} = \hat{W}^{R\, \mu}_{\alpha'\gamma} + i \hat{W}^{I\, \mu}_{\alpha'\gamma} \ ,
%\label{eq:PF17a-1}
%\end{equation}
%%
%where
%%
%\begin{align}
%\hat{W}^{R\, \mu}_{\alpha'\gamma} & \equiv W^\mu_{\alpha'\gamma}  \nonumber \\
%&  - \sum_{\beta,\beta' \ne \alpha'} W_{\alpha'\beta'} \left(G^IW \frac{1}{1 + (G^IW)^2}\right)_{\beta'\beta} G^I_\beta W^\mu_{\beta\gamma} \ , \nonumber \\
%\hat{W}^{I\, \mu}_{\alpha'\gamma} & \equiv  - \sum_{\beta,\beta'\ne\alpha'} W_{\alpha'\beta'} \left(\frac{1}{1 + (G^IW)^2}\right)_{\beta'\beta} G^I_\beta W^\mu_{\beta\gamma} \ .
%\nonumber \\
%\label{eq:PF17b-1}
%\end{align}
%%

The dressed photon vertex $\bra{\hat{F}^\mu_{Kr}}_\gamma$  in Eq.~(\ref{eq:PF28}) follows from Eq.~(\ref{eq:F25}):
\begin{equation}
\bra{\hat{F}^\mu_{Kr}}_\gamma \equiv \bra{F^\mu_{Kr}}_\gamma  - i \sum_\beta \bra{F_{Kr}}_\beta G^I_{\beta} \hat{\bar{W}}^\mu_{\beta\gamma} \ ,
\label{eq:PF25}
\end{equation}
\\
where, from Eq.~(\ref{eq:F10}),
\begin{align}
\bra{F^\mu_{Kr}}_\gamma & \equiv \bra{F^\mu_{0r}}_\gamma + \sum_\beta \bra{F_{0r}}_\beta G^R_\beta W^\mu_{\beta\gamma}  \nonumber \\
&  = \bra{F^\mu_{0r}}_\gamma + \sum_\beta \bra{F_{Kr}}_\beta G^R_\beta U^\mu_{\beta\gamma} \ ,
\label{eq:PF13}
\end{align}
and, from Eq.~(\ref{eq:F20a}), 
\begin{equation}
\hat{\bar{W}}^\mu_{\alpha'\gamma}  \equiv W^\mu_{\alpha'\gamma}  - i \sum_{\beta,\beta'} W_{\alpha'\beta'}  G^I_{\beta'} \left(\left(\bar{D}^X\right)^{-1}\right)_{\beta'\beta} W^\mu_{\beta\gamma} \ .
\label{eq:PF20a}
\end{equation}
%
%The above equation can be also rewritten as (cf.~ Eqs.~(\ref{eq:F20b},\ref{eq:F20c}))
%%
%\begin{equation}
%\hat{\bar{W}}^\mu_{\alpha'\gamma} = \hat{\bar{W}}^{\cal A\, \mu}_{\alpha'\gamma} + i \hat{\bar{W}}^{\cal B\, \mu}_{\alpha'\gamma}  \ ,
%\label{eq:PF20b}
%\end{equation}
%%
%with
%%
%\begin{align}
%\hat{\bar{W}}^{\cal A\, \mu}_{\alpha'\gamma} & \equiv W^\mu_{\alpha'\gamma}  \nonumber \\
%&  - \sum_{\beta,\beta'} W_{\alpha'\beta'} G^I_{\beta'}\left( \frac{1}{1 + (WG^I)^2} WG^I\right)_{\beta'\beta} W^\mu_{\beta\gamma} \ , \nonumber \\
%\hat{\bar{W}}^{\cal B\,\mu}_{\alpha'\gamma} & \equiv  - \sum_{\beta,\beta'} W_{\alpha'\beta'} G^I_{\beta'}\left(\frac{1}{1 + (WG^I)^2}\right)_{\beta'\beta} W^\mu_{\beta\gamma} \ .
%\nonumber \\
%\label{eq:PF20c}
%\end{align}
%%

\vskip 0.5cm
It is straightforward to show that Eq.~(\ref{eq:PF28}) reduces (as it should) to Watson's theorem for photoproduction below the first inelastic threshold \cite{Watson54}. To this end, we realize that the first term on the right-hand side of Eq.~(\ref{eq:PF28}) is the pole part of the photoproduction amplitude given by Eq.~(\ref{eq:H10}) as shown in Appendix~\ref{ap:C}. This equation, in turn, can be recast in terms of the pole and nonpole $K$-matrices [cf.~Eq.~(\ref{eq:PF10})] as $M^{P\, \mu} = N^X K^{P\, \mu}  - i T^P G^I (K^{P\, \mu} + W^\mu)$ through the substitutions $G \to -iG^I$, $V^{P\, \mu} \to K^{P\, \mu}$ and $V^{NP\, \mu} \to K^{NP\, \mu}\ (\equiv W^\mu)$. Then, below the first inelastic threshold, we have 
\begin{widetext}
\begin{align}
M^\mu_{\alpha'\gamma} & = \sum_{r'r} \left\{ N^X_{\alpha'}\ket{F_{Kr'}}_{\alpha'}  \left(\frac{1}{(E - m_{0})I - \Sigma_K + i  \bra{F_K}_{\alpha'} G^I_{\alpha'} N^X_{\alpha'}\ket{F_K}_{\alpha'}   }\right)_{r'r}
 \bra{\hat{F}^\mu_{Kr}}_\gamma \right\}  + N^X_{\alpha'} W^\mu_{\alpha'\gamma} \nonumber \\
& = N^X_{\alpha'} K^{P\, \mu}_{\alpha'\gamma}  - iT^P_{\alpha'\alpha'} G^I_{\alpha'} \left(K^{P\, \mu}_{\alpha'\gamma} + W^\mu_{\alpha'\gamma} \right)  + N^X_{\alpha'}W^\mu_{\alpha'\gamma} \nonumber \\
&  = \left( N^X_{\alpha'} - iT^P_{\alpha'\alpha'}G^I_{\alpha'}\right)\left(K^{P\, \mu}_{\alpha'\gamma} + W^\mu_{\alpha'\gamma}\right) = \left(1 - iT_{\alpha'\alpha'}G^I_{\alpha'}\right) K^\mu_{\alpha'\gamma}  = N_{\alpha'}K^\mu_{\alpha'\gamma} \nonumber \\
& = e^{\delta_{\alpha'}}\cos\delta_{\alpha'} K^\mu_{\alpha'\gamma} \ ,
\label{eq:PF20}
\end{align}
\end{widetext}
where we have also made use of Eqs.~(\ref{eq:PF17}) and (\ref{eq:PF25}) and of Eq.~(\ref{eq:F7}) for photoproduction, i.e., 
\begin{equation}
 K^\mu_{\alpha'\gamma} = K^{P\, \mu}_{\alpha'\gamma} + W^\mu_{\alpha'\gamma}
 \label{eq:PF10}
 \end{equation}
 with $W^\mu_{\alpha'\gamma}$ given by Eq.~(\ref{eq:PF11}) and
 \begin{equation}
 K^{P\, \mu}_{\alpha'\gamma} = \sum_{r'r} \ket{F_{Kr'}}_{\alpha'} S_{K r'r} \bra{F^\mu_{Kr}}_\gamma \ .
 \label{eq:PF12}
 \end{equation}
 % 
%Recall that $K^\mu_{\alpha'\gamma}$ is Hermitian if the driving term $V^\mu_{\alpha'\gamma}$ in Eq.~(\ref{eq:F7}) is.

\vskip 0.5cm

Equation (\ref{eq:PF28}) is the main result of this section. Together with Eq.~(\ref{eq:F28}) of the previous section, they may be used as the starting points in the construction of unitary isobar models. This is done in the following sections.

\vskip 0.5cm
Before leaving this section, a remark is in order. It is straightforward to show that if we use the form of  the nonpole $T$-matrix given by Eq.~(\ref{eq:F16}) for photoproduction,
\begin{equation}
X^\mu_{\alpha'\gamma}  = N^X_{\alpha'}\hat{W}^\mu_{\alpha'\gamma} \ ,
\label{eq:PF16-1}
\end{equation}
instead of that given by Eq.~(\ref{eq:F20}), the full dressed photoproduction vertex $\bra{F^\mu_r}_\gamma$  can be expressed in the form [cf. Eq.~(\ref{eq:F14})]
\begin{align}
\bra{F^\mu_r}_\gamma & \equiv \bra{F^\mu_{Kr}}_\gamma  - i \sum_{\beta} \bra{F_{Kr}}_\beta G^I_\beta X^\mu_{\beta\gamma} \nonumber \\
& = \bra{F^\mu_{Kr}}_\gamma  - i \sum_{\beta} \bra{F_{Kr}}_\beta G^I_\beta N^X_\beta \hat{W}^\mu_{\beta\gamma} \ ,
\label{eq:PF18}
\end{align}
instead of that form given by Eq.~(\ref{eq:F24}). Thus, one can replace the photon vertex $\bra{F^\mu_r}_\gamma = \bra{\hat{F}^\mu_{Kr}}_\gamma  N^X_\gamma = \bra{\hat{F}^\mu_{Kr}}_\gamma$ appearing in Eq.~(\ref{eq:PF28}) by 
the form given in the above equation.
Which of the two forms to use depends on what one wants to do. In the full calculation, where the channel couplings are fully taken into account, the form
given by Eq.~(\ref{eq:PF18}) would be preferable numerically, for it involves $\hat{W}^\mu_{\beta\gamma}$ which requires the matrix inversion of the same $D^X$ that enters in the calculation of the final state hadronic interaction part. In contrast, 
$\bra{F^\mu_r}_\gamma = \bra{\hat{F}^\mu_{Kr}}_\gamma$ involves $\hat{\bar{W}}^\mu_{\beta\gamma}$ that requires an independent matrix inversion of $\bar{D}^X$ from that for the final hadronic interaction.
In an approximate calculation, however, as discussed in the following sections, the form $\bra{F^\mu_r}_\gamma = \bra{\hat{F}^\mu_{Kr}}_\gamma$ may be more suitable.

\section{Possible approximations}\label{sec:UAIM}

The basic result of Sec.~\ref{sec:PS2} given by Eq.~(\ref{eq:F28}) and of Sec.~\ref{sec:PS3} given by Eq.~(\ref{eq:PF28}) provide a convenient starting point for possible approximations one can make 
with different levels of sophistications. In Eq.~(\ref{eq:F28}), the three basic ingredients for possible approximations are the nonpole $K$-matrix amplitude $W (\equiv K^{NP})$ as given by the integral equation (\ref{eq:F8a}), the dressed $K$-matrix resonance vertex $\ket{F_{Kr'}}$ ($\bra{F_{Kr}}$) given by Eq.~(\ref{eq:F10}) and the $K$-matrix self-energy $\Sigma_{K r'r}$ given by Eq.~(\ref{eq:F12}). These involve an integration over the loop momentum through the real part of the meson-baryon propagator $G^R_\beta$. Note that all the ingredients, the Watson's factor $N^X$, the dressed $K$-matrix vertex as well as the $K$-matrix self-energy, entering in Eq.~(\ref{eq:F28}) are expressed in terms of $W$. $W$ enters the $K$-matrix self-energy through the dressed $K$-matrix vertex. The different approximations one makes on the basic three ingredients just mentioned may be classified into few broad categories:
\begin{itemize}

\item[a)]
\textit{Unitary and Analytic Isobar Model} (UAIM) : In this approach, the driving nonpole term $U$ in Eq.~(\ref{eq:F8a}) is approximated by a phenomenological separable potential (see, e.g., Ref.~\cite{NBL90}) whose form allows to solve the integral equation for $W$ in Eq.~(\ref{eq:F8a}) analytically. The bare vertex $\ket{F_{0r'}}$( $\bra{F_{0r}}$) is obtained either from a microscopic Lagrangian or simply parametrized  phenomenologically.
 Then, the momentum-loop integration in Eq.~(\ref{eq:F10}) is carried out analytically to obtain $\ket{F_{Kr'}}$($\bra{F_{Kr}}$). $\Sigma_{K r'r}$ is obtained as given by Eq.~(\ref{eq:F12}), also by performing the momentum loop integration analytically. This model maintains unitarity and analyticity; the latter, by keeping explicitly both the real and imaginary parts of the meson-baryon propagator. Of course, the adopted separable potential should be analytic. Note that the contribution due to the real part of the meson-baryon propagator may lead to pole structures in the resulting reaction amplitude in the complex-energy plane that would correspond to dynamically generated resonances \cite{KSW95,OR10,BMM11}.

\item[b)]
\textit{Unitary Isobar Model} (UIM) : 
Here,  $W$ and $\ket{F_{Kr'}}$($\bra{F_{Kr}}$)  are directly parametrized in a completely phenomenological or semi-phenomenological manner, thereby avoiding to solve the integral equation for $W$ and the momentum-loop integration for $\ket{F_{Kr'}}$($\bra{F_{Kr}}$). Here, the self-energy $\Sigma_{K r'r}$ [cf. Eq.~(\ref{eq:F12})] is also simply parametrized. In this model, the analyticity of the original reaction amplitude is lost, because the momentum-loop integrations involving the real part of the meson-baryon propagator in Eqs.~(\ref{eq:F10}), (\ref{eq:F12}), and (\ref{eq:F8a}) are not performed. In general, ignoring the contributions arising from the real part of the meson-baryon propagator violates analyticity,
since the dispersion relation condition due to analyticity between the real and imaginary parts of the reaction amplitude  \cite{Ga66} will no longer be satisfied.

\end{itemize}

\section{Summary} \label{sec:SUM}

We have exposed the full complex phase structure of the meson-baryon $T$-matrix reaction amplitude in the coupled channels framework. By exhibiting the complex phase structure of the pole and nonpole parts of the $T$-matrix, we have achieved to express the reaction amplitude in a form which suitably serves as a starting point  for making approximations of varying degrees of sophistication.  In particular, it allows for approximations where the basic properties of the $S$-matrix, namely, unitarity and analyticity, can be maintained automatically. 
Recall that in earlier works \cite{O74a,DHKT99,DKT07,TDKV11} unitarity of the reaction amplitude in isobar models is implemented by imposing the unitarity condition on the resonance amplitude [pole amplitude $T^P$], separately from the unitarity condition on the background amplitude [nonpole amplitude $X (\equiv T^{NP})$]. 
In the present work no such additional condition is required. Here, the unitarity of $T^P$ arises automatically from the dressing mechanism inherent in the basic scattering equation (Eq.~(\ref{eq:A1})). 
In the case of photoproduction, gauge invariance can be satisfied as well. Furthermore, we have shown how the analog in meson-baryon reaction of the well-known Watson's theorem in photoproduction emerges in the present formulation. 

Finally, we mention that calculations based on a coupled channels unitary isobar model as described briefly in Sec.~\ref{sec:UAIM} will be reported shortly.

\begin{acknowledgments}
The authors thank Helmut Haberzettl for sharing his private note on the resonance coupling propagator. 
\end{acknowledgments}

\appendix

\section{Phase-shift parametrization} \label{ap:PSparam}

In this appendix, we give the phase-shift parametrization of the Watson's factors $N_\alpha$ and $\bar{N}_\alpha$ defined in  Eqs.~ (\ref{eq:A9}) and (\ref{eq:14}) as well as of the on-shell $\hat{K}_{\alpha\alpha}$ defined in Eq.~(\ref{eq:Khat}). Here, we confine ourselves to stable particles only and consider the channels whose on-shell 
elastic scattering $T$-matrix amplitude in partial-wave basis can be parametrize in terms of the phase-shift ($\delta_\alpha$) and
inelasticity ($\eta_\alpha$) as
\begin{equation}
\rho_\alpha T_{\alpha\alpha}  = \frac{i}{2} \left(\eta_\alpha e^{i2\delta_\alpha} - 1\right) \ ,
\label{eq:A16}
\end{equation}
with $\rho_\alpha$ denoting the (phase-space) density of state in the channel specified by the index $\alpha$.
%(cf. Eq.~(\ref{eq:B7})).
Then, inserting Eq.~(\ref{eq:A16}) into Eqs.~(\ref{eq:A9}) and (\ref{eq:14}), we have for the Watson factor\footnote{Note that, for a stable particles channel $\beta$, $G^I_{\beta} =  \pi \delta(E - H_{0\beta}) \to \rho_{\beta}$ after the momentum loop integration.}% (cf. Eq.~(\ref{eq:B7})).}
\begin{equation}
N_{\alpha}  = \bar{N}_\alpha = 1 -i\rho_{\alpha} T_{\alpha\alpha} = \frac12 \left(\eta_\alpha e^{i2\delta_\alpha} + 1\right)  \ . 
\label{eq:A11}
\end{equation}

Inserting the above two equations into Eq.~(\ref{eq:A8}), and solving for $\hat{K}_{\alpha\alpha}$, we obtain
\begin{equation}
\hat{K}_{\alpha\alpha} = - \frac{1}{\rho_\alpha} ~ \frac{2\eta_\alpha\sin2\delta_\alpha + i(1 - \eta_\alpha^2)}{1 + \eta_\alpha^2 + 2\eta_\alpha\cos2\delta_\alpha}\ .
\label{eq:D2}
\end{equation}

This result reveals a very simple phase structure of the on-shell $\hat{K}_{\alpha\alpha}$ in terms of the phase-shift and inelasticity of the elastic scattering $T$-matrix amplitude.

Below the inelastic threshold ($\eta_\alpha=1$), Eq.~(\ref{eq:D2}) reduces to
\begin{equation}
\hat{K}_{\alpha\alpha} = K_{\alpha\alpha} = - \frac{\sin2\delta_\alpha}{\rho_\alpha\left(1 + \cos2\delta_\alpha\right)} =
-\frac{1}{\rho_\alpha}\tan\delta_\alpha \ ,
\label{eq:D3}
\end{equation}
and, as it should, one recovers the phase-shift parametrization of the on-shell $K$-matrix $K_{\alpha\alpha}$ (valid even above the inelastic threshold) which is a pure real quantity.  
%Note that, here, the above result is a consequence of phase-shift parametrization of the on-shell $T$-matrix, which, in turn, is a consequence of unitarity and, therefore, $V$ should be Hermitian.  The above result also corroborates Eq.~(\ref{eq:C0}).

Inserting the phase-shift parametrization of the on-shell $K$-matrix into Eq.~(\ref{eq:A17a}), yields
\begin{equation}
N_{K\, \alpha} = \bar{N}_{K\, \alpha} = e^{i\delta_\alpha} \cos\delta_\alpha \ .
\label{eq:D3a}
\end{equation}

\vskip 0.5cm
In complete analogy to the phase-shift parametrization of the on-shell elastic $T$-matrix amplitude [cf. Eq.~(\ref{eq:A16})], if we assume the corresponding phase-shift parametrization of the on-shell elastic nonpole $T$-matrix ($X \equiv T^{NP}$) in Eq.~(\ref{eq:F2}) to be
\begin{equation}
\rho_\alpha X_{\alpha\alpha}  = \frac{i}{2} \left(\eta^X_\alpha e^{i2\delta^X_\alpha} - 1\right) \ ,
\label{eq:D4}
\end{equation}
then, the corresponding Watson's factors $N^X_\alpha$ and $\bar{N}^X_\alpha$ defined by Eqs.~(\ref{eq:F18}) and (\ref{eq:F21}) become
\begin{equation}
N^X_{\alpha}  = 1 -i\rho_{\alpha} X_{\alpha\alpha} = \frac12 \left(\eta^X_\alpha e^{i2\delta^X_\alpha} + 1\right) = \bar{N}^X_\alpha \ . 
\label{eq:D5}
\end{equation}

For the on-shell $\hat{W}_{\alpha\alpha}$, we obtain
\begin{equation}
\hat{W}_{\alpha\alpha} = - \frac{1}{\rho_\alpha} ~ \frac{2\eta^X_\alpha\sin2\delta^X_\alpha + i(1 - \eta^X_\alpha{^2})}{1 + \eta^X_\alpha{^2} + 2\eta^X_\alpha\cos2\delta^X_\alpha} \ ,
\label{eq:D6}
\end{equation}

and below the inelastic threshold ($\eta^X_\alpha = 1$), it reduces to 
\begin{equation}
\hat{W}_{\alpha\alpha} = W_{\alpha\alpha} = - \frac{\sin2\delta^X_\alpha}{\rho_\alpha\left(1 + \cos2\delta^X_\alpha\right)} =
-\frac{1}{\rho_\alpha}\tan\delta^X_\alpha \ .
\label{eq:D7}
\end{equation}

\section{Pole and nonpole decomposition of the $T$-matrix reaction amplitude} \label{ap:C}

Although the pole and nonpole decomposition of the meson-baryon $T$-matrix reaction amplitude is widely used in the literature (see, e.g., \cite{AS81,MSL07}), due to its central role in the present work, its derivation is provided in this Appendix.
We will also decompose the photoproduction amplitude starting from the gauge-invariant amplitude obtained from the field theoretic considerations \cite{H97}. 
In this Appendix, the reference to two-particle channels are suppressed for the sake of not overloading with unessential notations in  the derivation.

\subsection{Meson-baryon $T$-matrix reaction amplitude}

The meson-baryon $T$-matrix obeys the Lippmann-Schwiger-type scattering equation
\begin{equation}
T = V + VGT \ .
\label{eq:C1}
\end{equation}
It can be recast into the form
\begin{equation}
T = T^P + T^{NP} \ ,
\label{eq:C2}
\end{equation}
with 
\begin{equation}
T^{NP} = V^{NP} + V^{NP} G T^{NP} \ , 
\label{eq:C3}
\end{equation}
where $V^{NP}$ stands for one-nucleon irreducible potential (the nonpole part 
of $V$), i.e.,
\begin{equation}
V^{NP} = V - V^P \ ,
\label{eq:C4}
\end{equation}
with the one-nucleon reducible potential $V^P$ (the pole part of $V$) given by
\begin{equation}
V^P = \sum_r \ket{F_{0r}}S_{0r}\bra{F_{0r}} \ .
\label{eq:C5}
\end{equation}
In the above equation, $\ket{F_{0r}}$ denotes the bare vertex and $S_{0r}$, 
the bare baryon propagator for a given bare resonance $r$, including the nucleon
($r=N$). 

Below, we show how the pole T-matrix, $T^P$, in Eq.~(\ref{eq:C2}) can be expressed 
in a compact form. For this purpose, let us start from 
Eqs.~(\ref{eq:C1}) to (\ref{eq:C4}) to express $T^P$ as
\begin{eqnarray}
T^P & = & \left(1 + T^{NP}G\right) V^P +  T^P G V \nonumber \\
T^P(1 - GV) & = & \left(1 + T^{NP}G\right) V^P \nonumber \\
T^P & = & \left(1 + T^{NP}G\right) V^P (1 -GV)^{-1} \nonumber \\
&     = & \left(1 + T^{NP}G\right) V^P (1 + GT)    \nonumber \\
&     = &  \left(1 + T^{NP}G\right) V^P\left[\left(1 + GT^{NP}\right) + GT^P\right] \ . \nonumber \\
\label{eq:C6}
\end{eqnarray}
Inserting Eq.~(\ref{eq:C5}) into Eq.~(\ref{eq:C6}), we have
\begin{equation}
T^P = \sum_r \Big\{ \ket{F_r} S_{0 r} \bra{F_r} 
    + \ket{F_r} S_{0 r} \bra{F_{0 r}} G T^P \Big\} \ ,
\label{eq:C7}
\end{equation}
with the dressed vertex defined as
\begin{align}
\ket{F_r} & \equiv \left(1 + T^{NP}G\right)\ket{F_{0 r}} \ , \nonumber \\
\bra{F_r} & \equiv \bra{F_{0 r}}\left(1 + G T^{NP}\right) \ , \nonumber \\
 \label{eq:C8}
\end{align}

Multiplication of Eq.~(\ref{eq:C7}) by $\bra{F_{0 r'}}G$ from the left gives
\begin{eqnarray}
\bra{F_{0 r'}}G T^P & = & \sum_r \Big\{\Sigma_{r'r} S_{0 r} \bra{F_r} 
    + \Sigma_{r'r} S_{0 r} \bra{F_{0 r}} G T^P \Big\} \nonumber \\
 \sum_r \Sigma_{r'r}S_{0 r}&\bra{F_r} & =  \sum_r\left( S^{-1}_{0 r}\delta_{r'r} - \Sigma_{r'r}\right) S_{0 r} \bra{F_{0 r}} G T^P \ , \nonumber \\
\label{eq:C9}
\end{eqnarray}
where the last equality is a simple rearrangement of the first equality together with the introduction of the self-energy matrix
\begin{equation}
\Sigma_{r'r} \equiv \bra{F_{0 r'}}G \ket{F_r} \ .
\label{eq:C10}
\end{equation}

Defining the dressed propagator matrix
\begin{equation}
S^{-1}_{r'r} \equiv S^{-1}_{0 r}\delta_{r'r} - \Sigma_{r'r} \ ,
\label{eq:C11}
\end{equation}
we have, from Eq.~(\ref{eq:C9}),
\begin{eqnarray}
 \sum_{r'r}S_{kr'} \Sigma_{r'r} S_{0 r} \bra{F_r} & = & \sum_{r'r}S_{kr'}S^{-1}_{r'r} S_{0 r} \bra{F_{0 r}} G T^P \nonumber \\
%& = & \sum_{r}\delta_{kr} S_{o r} \bra{F_{o r}} G T^P \nonumber \\
%& = & \sum_{r'r}S_{kr'}\left(S^{-1}_{o r'}\delta_{r'r} - S^{-1}_{r'r}\right) 
%S_{o r} \bra{F_r} \nonumber \\
& = &  S_{0 k} \bra{F_{0 k}} G T^P \ .
%& = & \sum_{r'r}S_{kr'}\left(S^{-1}_{o r'}\delta_{r'r} - S^{-1}_{r'r}\right)
%S_{o r} \bra{F_r}
\label{eq:C12}
\end{eqnarray}

Inserting the above result back into Eq.~(\ref{eq:C7}), we have
\begin{eqnarray}
T^P & = & \sum_r \Big\{ \ket{F_r} S_{0 r} \bra{F_r} 
    + \ket{F_r} \sum_{r'l}S_{rl} \Sigma_{lr'} S_{0 r'} \bra{F_{r'}}\Big\} \nonumber \\
%    & = & \sum_{rr'} \ket{F_r} \Big\{ \delta_{rr'}S_{o r'} 
%    + \sum_l S_{rl} \Sigma_{lr'} S_{o r'} \Big\}\bra{F_{r'}} \nonumber \\
    & = & \sum_{rr'} \ket{F_r} \Big\{ \delta_{rr'}S_{0 r'}  \nonumber \\
& & \qquad\qquad   + \sum_l S_{rl} \left(S^{-1}_{0 l}\delta_{lr'} - S^{-1}_{lr'}\right)S_{0 r'}
       \Big\} \bra{F_{r'}} \nonumber \\
%    & = & \sum_{rr'} \ket{F_r} \Big\{ \delta_{rr'} S_{o r'}
%      + S_{rr'} - \delta_{rr'} S_{o r'} \Big\} \bra{F_{r'}} \nonumber \\
    & = & \sum_{rr'} \ket{F_r} S_{rr'} \bra{F_{r'}} \ . \nonumber \\
\label{eq:C13}
\end{eqnarray}

\subsection{Photoproduction reaction amplitude}

Following the field theoretic approach of Haberzettl \cite{H97}, the gauge-invariant photoproduction 
amplitude in the one-photon approximation can be expressed as  
\begin{equation}
M^\mu = V^\mu  + TGV^\mu  ,
\label{eq:H1}
\end{equation}
with $\mu$ denoting the Lorentz index of the photon polarization and 
\begin{equation}
V^\mu = \tilde{m}^\mu_s + M^\mu_u + M^\mu_t + m^\mu_{KR} + U^\mu G\ket{F_N} \ , 
\label{eq:H2}
\end{equation}
where $U^\mu$ stands for the exchange current which arises from the coupling of the 
photon to $V^{NP}$, $m^\mu_{KR}$ stands for the sum of the bare Kroll-Ruderman contact 
current arising from the direct coupling of the photon to the bare vertex 
$\ket{F_{0r}}$ appearing in Eq.~(\ref{eq:C5}), $M^\mu_x (x = u, t)$ denotes the 
$x$-channel (tree-level Feynman diagram) contribution
%%
%\begin{align}
%M^\mu_u & = \sum_{rr'} \bra{F^\mu_r}S_{rr'}\ket{F_{r'}} \ , \nonumber \\
%M^\mu_t & = \bra{F_N}\Delta_\pi\bra{F^\mu_\pi} \ ,
%\label{eq:H3a}
%\end{align}
%%
and, $\tilde{m}^\mu_s$ stands 
for the $s$-channel bare current. The latter is given by
\begin{equation}
\tilde{m}^\mu_s = \sum_r \ket{F_{0r}}S_{0r}\bra{\tilde{F}^\mu_{0r}} \ ,
\label{eq:H3}
\end{equation}
where $\bra{\tilde{F}^\mu_{0r}}$ is defined as
\begin{equation}
\bra{\tilde{F}^\mu_{0r}} \equiv \bra{F^\mu_{0r}} +  \bar{m}^\mu_{KR r}G\ket{F_N} \ ,
\label{eq:H4}
\end{equation}
with $\bra{F^\mu_{0r}}$ denoting the bare $rN\gamma$ vertex and, $\bar{m}^\mu_{KR r}$,
the bare Kroll-Ruderman term for a given resonance $r$ in $m^\mu_{KR}$ with the meson leg 
reversed, i.e., $NM\gamma \to r$.

Now, analogous to Eq.~(\ref{eq:C4}), we decompose $V^\mu$ in Eq.~(\ref{eq:H2}) as 
\begin{equation}
V^\mu  =  V^{P \mu} + V^{NP \mu} \ ,
\label{eq:H5}
\end{equation}
where
\begin{equation}
V^{P \mu} \equiv  \tilde{m}^\mu_s = \sum_r \ket{F_{0r}}S_{0r}\bra{\tilde{F}^\mu_{0r}} \ ,
\label{eq:H6}
\end{equation}
and 
\begin{equation}
V^{NP \mu} \equiv M^\mu_u + M^\mu_t + m^\mu_{KR} + U^\mu G\ket{F_N} \ . 
\label{eq:H7}
\end{equation}

Inserting Eqs.~(\ref{eq:C2}) and (\ref{eq:H5}) into Eq.~(\ref{eq:H1}), we have
\begin{eqnarray}
M^\mu & = & V^{P \mu} + V^{NP \mu} + (T^P+T^{NP})G( V^{P \mu}  + V^{NP \mu}) \nonumber \\ 
      & = & M^{P \mu} +  M^{NP \mu} \ ,
\label{eq:H8}
\end{eqnarray}
where
\begin{equation}
M^{NP \mu} \equiv V^{NP \mu} + T^{NP}GV^{NP \mu} \ ,
\label{eq:H9}
\end{equation}
and
\begin{equation}
M^{P \mu} \equiv V^{P \mu} + T^{NP}GV^{P \mu} +  T^PG( V^{P \mu}  + V^{NP \mu}) \ .
\label{eq:H10}
\end{equation}

Now, using Eqs.~(\ref{eq:C8}) and (\ref{eq:H6}), 
\begin{eqnarray}
V^{P \mu} + T^{NP}GV^{P \mu} 
%& = & (1 + T^{NP}G)V^{P \mu} \nonumber \\
%& = & \sum_r (1 + T^{NP}G)\ket{F_{o r}} S_{o r} \bra{\tilde{F}^\mu_{o r}} \nonumber \\
& = & \sum_r \ket{F_r} S_{0 r}\bra{\tilde{F}^\mu_{0 r}} \ .
\label{eq:H11}
\end{eqnarray}

Similarly, using Eqs.~(\ref{eq:C10}), (\ref{eq:C13}), and (\ref{eq:H6}),
\begin{eqnarray}
T^PG&(&V^{P \mu}  + V^{NP \mu})  = \nonumber \\
%& = & \sum_{rr'}\ket{F_r} S_{rr'} \bra{F_{r'}}G( V^{P \mu} + V^{NP \mu}) \nonumber \\
& = & \sum_{rr'} \ket{F_r} S_{rr'} \Big[\bra{F_{r'}}G \sum_{l}\ket{F_{0 l}} S_{0 l} 
\bra{\tilde{F}^\mu_{0 l}}  \nonumber \\
& & \ \ \qquad\qquad\qquad\qquad\qquad\qquad + \bra{F_{r'}}G V^{NP \mu} \Big] \nonumber \\
& = & \sum_{rr'l}\ket{F_r} S_{rr'} \left[\Sigma_{r'l} S_{0 l} \bra{\tilde{F}^\mu_{0 l}}
  + \delta_{r'l}\bra{F_l}G V^{NP \mu} \right] \ . \nonumber \\
\label{eq:H12}
\end{eqnarray}

Inserting Eqs.~(\ref{eq:H11}) and (\ref{eq:H12}) into Eq.~(\ref{eq:H10}), and with help of Eq.~(\ref{eq:C11}), 
we have
\begin{eqnarray}
M^{P \mu} %& = & \sum_r \ket{F_r} S_{o r} \bra{\tilde{F}^\mu_{o r}} \nonumber \\
%& &  + \sum_{rr'l} \ket{F_r} S_{rr'} \left[\Sigma_{r'l} S_{o l} \bra{\tilde{F}^\mu_{o l}}
% + \delta_{r'l}\bra{F_l}G V^{NP \mu} \right] \nonumber \\
& = & \sum_{rr'l}\ket{F_r}S_{rr'} \Big[ (S)^{-1}_{r'l}S_{0 l}\bra{\tilde{F}^\mu_{0 l}}
+ \Sigma_{r'l} S_{0 l} \bra{\tilde{F}^\mu_{0 l}}  \nonumber \\
&& \qquad\qquad\qquad\qquad\qquad\qquad + \delta_{r'l} \bra{F_l}G V^{NP \mu} 
\Big] \nonumber \\
& = & \sum_{rr'l} \ket{F_r}S_{rr'} \Big[ \left\{(S)^{-1}_{r'l} + \Sigma_{r'l} \right\} 
S_{0 l} \bra{\tilde{F}^\mu_{0 l}} \nonumber \\
& &\qquad\qquad\qquad\qquad\qquad\qquad + \delta_{r'l}\bra{F_l}G V^{NP \mu} \Big] \nonumber \\
%& = & \sum_{rr'l} \ket{F_r}S_{rr'} \Big[ \delta_{r'l}(S_o)^{-1}_lS_{o l} 
%\bra{\tilde{F}^\mu_{o l}}  \nonumber \\
%& &\qquad\qquad \qquad\qquad\qquad\qquad + \delta_{r'l}\bra{F_l}G V^{NP \mu} \Big] \nonumber \\
%& = & \sum_{rr'} \ket{F_r}S_{rr'} \left[ \bra{\tilde{F}^\mu_{o r'}} + 
%\bra{F_{r'}}G V^{NP \mu} \right] \nonumber \\
& = & \sum_{rr'} \ket{F_r}S_{rr'}\bra{F^{\mu}_{r'}}  \ ,
\label{eq:H13}
\end{eqnarray}
where, in the last equality above, we have introduced the dressed electromagnetic vertex 
\begin{align}
\bra{F^{\mu}_{r'}} & \equiv \bra{\tilde{F}^\mu_{0 r'}} + \bra{F_{r'}}G V^{NP \mu} \nonumber \\
& = \bra{\tilde{F}^\mu_{0r'}} + \bra{F_{0r'}}G M^{NP \mu} \ .
\label{eq:H14}
\end{align}

\section{Two-resonance coupling}\label{ap:TRC}

The resonance propagator appearing in the pole part of the $T$-matrix [cf. Eq.~(\ref{eq:F28})] is, in general, a matrix in resonance space. In most of the cases, there is only one resonance for a given partial-wave state, in which case, the propagator reduces to a number. In other cases, such as in the $\pi N$ $S_{11}$ partial wave,  there can be two resonances close to each other [$S_{11}(1535)$ and $S_{11}(1650)$] which causes a considerable resonance coupling effect. For the two-resonance case, the resonance propagator matrix $S$ can be obtained explicitly.  Following Ref.~\cite{HH},  we have,
\begin{equation}
S^{-1} = S^{-1}_0 - \Sigma  = 
\begin{pmatrix}
 E- m_{0\, 1} - \Sigma_{11}  & - \Sigma_{12}  \\  - \Sigma_{21} & E - m_{0\, 2} - \Sigma_{22}
  \end{pmatrix} \ ,
\label{eq:P2}
\end{equation}
and hence
\begin{align}
S & =  \left(\frac{1}{S^{-1}_0 - \Sigma}\right) \nonumber \\
& = \frac{1}{|D|}
\begin{pmatrix}
 E - m_{0\, 2} - \Sigma_{22} & \Sigma_{12} \\  \Sigma_{21} & E- m_{0\, 1} - \Sigma_{11}
  \end{pmatrix} \ ,
\label{eq:P3}
\end{align}
where $|D|$ stands for the determinant of $S^{-1}$, 
\begin{equation}
|D| =  (E- m_{0\, 1} - \Sigma_{11})(E- m_{0\, 2} - \Sigma_{22}) - \Sigma_{12}\Sigma_{21}  \ .
\label{eq:P4}
\end{equation}

Now, defining
\begin{equation}
\mu_i \equiv m_{0\, i} + \Sigma_{ii}  \ \ \ \ \ \ \rm{and} \ \ \ \ \ C^2 \equiv \Sigma_{12}\Sigma_{21} \ ,
\label{eq:P5}
\end{equation}
the pole condition reads
\begin{equation}
|D| = (E-\mu_1) (E- \mu_2) - C^2 = 0 \ ,
\label{eq:P7}
\end{equation}
producing two solutions
\begin{equation}
E\ \   \to \ \   M_\pm = \frac{\mu_1 + \mu_2}{2} \pm \frac12 \sqrt{ (\mu_1 - \mu_2)^2 + 4C^2} \ .
\label{eq:P8}
\end{equation}

We then have
\begin{align}
\frac{1}{|D|} & = \frac{1}{(E - M_-)(E - M_+)} \nonumber \\
& = \frac{1}{(E-M_-) + (E-M_+)} \left[ \frac{1}{E - M_-} + \frac{1}{E- M_+}\right] \nonumber \\
& = \frac{1}{(E-\mu_1) + (E-\mu_2)} \left[ \frac{1}{E - M_-} + \frac{1}{E- M_+}\right] \ ,
\label{eq:P9}
\end{align}
where the equality $M_- + M_+ = \mu_1 + \mu_2$ has been used.

Equation (\ref{eq:P9}) allows to write the propagator in the form
\begin{equation}
S = \frac{R}{E - M_-} + \frac{R}{E - M_+}  \ , 
\label{eq:P10}
\end{equation}
with the ``residue" matrix $R$  given by
\begin{equation}
R \equiv  
\begin{pmatrix}
 \frac{E - \mu_{2}}{d} & \frac{\Sigma_{12}}{d} \\  \frac{\Sigma_{21}}{d} & \frac{E- \mu_{1}}{d}
  \end{pmatrix} \ , \ \ \ \ \ \ \  \frac{1}{d} = \frac{1}{(E-\mu_1) + (E-\mu_2)} \ .
\label{eq:P11}
\end{equation}

\vskip 0.5cm
If the resonance coupling is small enough, i.e., $\Sigma_{12} \sim \Sigma_{21} \sim 0$, then $M_\pm$ and $R$ reduce to
\begin{equation}
M_+ = \mu_1   \ , \ \ \ \ \ M_- = \mu_2 \ , \ \ \ \ \ \ { \rm{and}} \ \ \ \ 
R =
\begin{pmatrix}
 \frac{E - \mu_{2}}{d} & 0 \\  0 & \frac{E- \mu_{1}}{d}
  \end{pmatrix} \ ,
\label{eq:P12}
\end{equation}
so  that
\begin{equation}
S = 
\begin{pmatrix}
 \frac{1}{E - \mu_1} & 0 \\  0 & \frac{1}{E- \mu_2}
  \end{pmatrix}    \ ,
\label{eq:P13}
\end{equation}
as it should be.

%\bibliography{bibliography}

%\end{document}

\end{document}